\documentclass{article}
\usepackage{spconf}

\onecolumn

\usepackage[
    hidelinks,
]{hyperref}

\usepackage{graphicx}

\usepackage[
    noend,
]{algpseudocode}

\usepackage[
    table,
    dvipsnames,
]{xcolor}

\usepackage{xspace}

\usepackage{cite}

\usepackage{relsize}

\usepackage{			
	amsmath,
	amssymb,
	amsfonts,
}

\usepackage{siunitx}

\usepackage{tikz}
\usetikzlibrary{
	positioning,				
	shapes.geometric,			
	decorations.pathreplacing,	
	shadows,					
	patterns,					
}

\usepackage[
	toc=True,
	numberline=False,
	style=tree,
	sort=standard,
	nopostdot=True,
	acronym,
]{glossaries}

\glsdisablehyper  
\setacronymstyle{long-sm-short}  

\newacronym{sota}{SotA}{State of the Art}
\newacronym
    [longplural={Degrees of Freedom}]
    {dof}{DoF}{Degree of Freedom}

\newacronym{3gpp}{3GPP}{3rd Generation Partnership Program}

\newacronym{embb}{eMBB}{Enhanced Mobile Broadband}
\newacronym{urllc}{URLLC}{Ultra Reliable and Low Latency Communications}
\newacronym{mmtc}{mMTC}{Massive Machine Type Communications}

\newacronym{tx}{Tx}{Transmit}
\newacronym{rx}{Rx}{Receive}

\newacronym
    [longplural={Channel State Information at Transmitter}]
    {csit}{CSIT}{Channel State Information at Transmitter}
\newacronym{ber}{BER}{Bit Error Rate}
\newacronym{snr}{SNR}{Signal to Noise Ratio}

\newacronym{sgd}{SGD}{Stochastic Gradient Descent}
\newacronym{mmse}{MMSE}{Minimum Mean Squared Error}
\newacronym{zf}{ZF}{Zero Forcing}

\newacronym{dtft}{DTFT}{Discrete Time Fourier Transform}
\newacronym{dft}{DFT}{Discrete Fourier Transform}
\newacronym{fft}{FFT}{Fast Fourier Transform}

\newacronym{ofdm}{OFDM}{Orthogonal Frequency Division Multiplex}
\newacronym{ofdma}{OFDMA}{Orthogonal Frequency Division Multiple Access}
\newacronym{noma}{NOMA}{Non Orthogonal Multiple Access}

\newacronym{siso}{SISO}{Single Input Single Output}
\newacronym{simo}{SIMO}{Single Input Multiple Output}
\newacronym{miso}{MISO}{Multiple Input Single Output}
\newacronym{mimo}{MIMO}{Multiple Input Multiple Output}

\newacronym{ml}{ML}{Machine Learning}
\newacronym{dl}{DL}{Deep Learning}
\newacronym{rl}{RL}{Reinforcement Learning}

\newacronym{nn}{NN}{Neural Network}
\newacronym{ann}{AcNN}{Actor Neural Network}
\newacronym{cnn}{CrNN}{Critic Neural Network}

\newacronym{dqn}{DQN}{Deep Q Network}
\newacronym{ddpg}{DDPG}{Deep Deterministic Policy Gradient}
\newacronym{sac}{SAC}{Soft Actor-Critic}
\newacronym{ppo}{PPO}{Proximal Policy Optimization}

\newacronym{ntn}{NTN}{Non Terrestrial Networks}
\newacronym{leo}{LEO}{Low Earth Orbit}
\newacronym{geo}{GEO}{Geostationary}

\newacronym{sic}{SIC}{Successive Interference Cancellation}
\newacronym{los}{LoS}{Line-of-Sight}
\newacronym
    [longplural={Angles of Departure}]
    {aod}{AoD}{Angle of Departure}
\newacronym{awgn}{AWGN}{Additive White Gaussian Noise}
\newacronym{aoa}{AoA}{Angle of Arrival}
\newacronym{ula}{ULA}{Uniform Linear Array}
\newacronym{rsma}{RSMA}{Rate-Splitting Multiple Access}
\newacronym{sdma}{SDMA}{Space-Division Multiple Access}
\newacronym{oma}{OMA}{Orthogonal Multiple Access}
\newacronym{mrt}{MRT}{Maximum Ratio Transmission}
\newacronym{iui}{IUI}{Inter-User Interference}
\newacronym{uav}{UAV}{Unmanned Aerial Vehicle}

\newglossaryentry{ex}{%
	name={example},
	description={an example},
}%

\title
{%
	Learning Model-Free Robust Precoding \\ for Cooperative Multibeam Satellite Communications%
}

\name{%
    \centering
    {%
        \setlength\tabcolsep{1cm}
        \begin{tabular}{ccc}
            Steffen~Gracla
            & Alea~Schröder
            & Maik~Röper
            \\
            Carsten~Bockelmann
            & Dirk~Wübben
            & Armin~Dekorsy
        \end{tabular}
    }
    \thanks{%
        This work was partly funded by the German Ministry of Education and Research (BMBF) under grant 16KIS1028 (MOMENTUM) and 16KISK016 (Open6GHub) and the European Space Agency (ESA) under contract number 4000139559/22/UK/AL (AIComS).
        Parts of this work was submitted to IEEE 2023 IEEE International Conference on Acoustics, Speech and Signal Processing.
    }%
}

\address{%
    Dept. of Communications Engineering, University of Bremen, Bremen, Germany\\
    {Email: \{gracla, schroeder, roeper, bockelmann, wuebben, dekorsy\}@ant.uni-bremen.de}
}

\input{99_custommacros.sty}
\newcommand{\numusers}{K}
\newcommand{\numsatellites}{M}
\newcommand{\numantennasper}{N}

\newcommand{\useridx}{k}
\newcommand{\satidx}{m}
\newcommand{\antidx}{n}

\newcommand{\overallphase}{\varphi}

\newcommand{\otheruseridx}{l}

\newcommand{\erroraod}{\varepsilon}
\newcommand{\errorbound}{\Delta \varepsilon}
\newcommand{\errorphase}{\zeta}
\newcommand{\errorphasevariance}{\sigma_\zeta^2}
\newcommand{\errorphasescale}{\sigma_\zeta}

\newcommand{\datasymbol}{s}

\newcommand{\transmitpower}{P}
\newcommand{\noisepower}{\sigma_n^2}

\newcommand{\antdist}{d_{\antidx}}
\newcommand{\dist}{d}
\newcommand{\meanuserdist}{\bar{D}_\text{Usr}}
\newcommand{\userdist}{{D}_\text{Usr}}

\newcommand{\usergain}{G_\text{Usr}}
\newcommand{\satgain}{G_\text{Sat}}

\newcommand{\wavelength}{\lambda}
\newcommand{\aod}{\nu}
\newcommand{\steeringvec}{\mathbf{v}}
\newcommand{\steeringentry}{v}

\newcommand{\csimatrix}{\mathbf{H}}
\newcommand{\csivector}{\mathbf{h}}
\newcommand{\precodingmatrix}{\mathbf{W}}
\newcommand{\precodingvec}{\mathbf{w}}
\newcommand{\precodingentry}{w}

\newcommand{\sumrate}{R}

\newcommand{\statevec}{\mathbf{\internalstate}}
\newcommand{\actionsca}{\internalaction}
\newcommand{\actionvec}{\mathbf{\internalaction}}
\newcommand{\actionindex}{i}
\newcommand{\reward}{\sumrate}
\newcommand{\rewardapprox}{\hat{\sumrate}}

\newcommand{\criticnetworksca}{\internalcritic}
\newcommand{\actornetworksca}{\internalactor}
\newcommand{\actornetworkmean}{\bar{\internalactor}}
\newcommand{\actornetworkscale}{\sigma}
\newcommand{\actornetworkvec}{\boldsymbol{\internalactor}}
\newcommand{\params}{\boldsymbol{\internalparameter}}
\newcommand{\paramsactor}{\params_{\internalactor}}
\newcommand{\paramscritic}[1]{\params_{\internalcritic_{#1}}}
\newcommand{\paramsactortime}{\params_{\internalactor, \timeindex}}
\newcommand{\paramscritictimegeneric}{\params_{\internalcritic, \timeindex}}
\newcommand{\paramscritictime}[1]{\params_{\internalcritic_{#1}, \timeindex}}

\newcommand{\loss}{\mathcal{L}}
\newcommand{\lossactor}{\loss_{\internalactor}}
\newcommand{\lossactorvalue}{\loss_{\internalactor, 1}}
\newcommand{\lossactorentropy}{\loss_{\internalactor, 2}}
\newcommand{\losscritic}{\loss_{\internalcritic}}

\newcommand{\logentropyscale}{\alpha}
\newcommand{\learningbatchsize}{B}

\begin{document}
    \ninept
	\maketitle%
	%
	

\begin{abstract}
	Direct \acrlong{leo} satellite-to-handheld links are expected to be part of a new era in satellite communications.
    \acrlong{sdma} precoding is a technique that reduces interference among satellite beams, therefore increasing spectral efficiency by allowing cooperating satellites to reuse frequency.
    Over the past decades, optimal precoding solutions with perfect channel state information have been proposed for several scenarios, whereas robust precoding with only imperfect channel state information has been mostly studied for simplified models. In particular, for \acrlong{leo} satellite applications such simplified models might not be accurate.
    In this paper, we use the function approximation capabilities of the \acrlong{sac} deep \acrlong{rl} algorithm to learn robust precoding with no knowledge of the system imperfections.
\end{abstract}


\glsresetall  
%
	

\section{Introduction}

Integrating \gls{ntn}, e.g. satellites and \glspl{uav}, into current terrestrial infrastructure is one of the important pillars in the development of the sixth-generation standard of mobile communication networks \cite{3GPP.TR.38.863}. So-called holistic 3D networks will enable ubiquitous global coverage, provide capacity for temporally and locally varying traffic demands and enhance the robustness of terrestrial network infrastructure \cite{Leyva-Mayorga2020, Qu}. However, a host of challenges is introduced due to the high dynamism of \gls{ntn} devices.
\gls{leo} satellites have especially fast-changing \gls{los} channels, which are mainly characterized by the relative positions between satellites and users.
To increase spectral efficiency through frequency reuse, precoding based \gls{sdma} is used in satellite communications~\cite{vazquez2018precoding}.
However, the performance suffers due to imperfect positional information~\cite{MaikBeamspace,liu2022robust}.

Finding a precoding algorithm that maximizes performance metrics such as the sum rate for multiple geometric constellations while also showing robustness against imperfect positioning and channel estimates can prove challenging. In light of this, \gls{ml} methods present themselves as an attractive choice. \gls{ml} can be used to approximate a viable algorithm where the optimum is either infeasible to determine or wholly unavailable. \gls{dl} in particular has demonstrated tremendous potential on such problems in the past decade, \cite{mnih2013playing, dahrouj_overview_2021}.
Applying \gls{dl} to precoding has recently started to gather more attention, primarily in terrestrial communications, \eg \cite{zhang_data_2022} use supervised learning to approximate a lower complexity \gls{mmse} precoder, \cite{lee_deep_2020} show the ability of \gls{rl} precoders to optimally learn on toy scenarios without interference and\cite{sohrabi_robust_2020} use an autoencoder structure to learn robust precoding and decoding under imperfect channel knowledge.
In the \gls{leo} satellite context, \cite{liu2022robust} have extended their work on robust precoding pertaining imperfect positional knowledge for a single satellite scenario by a supervised low-complexity approximation.
In this paper, we will 
use model-free deep \gls{rl}. In \gls{rl}, an agent probes the environment (\ie selecting a precoding matrix and observing the result), thereby generating data to learn from, to gain understanding of the system dynamics and adjust their behavior to maximize an objective.
By using this approach, no assumptions about the error modeling need to be made; it is instead discovered and inferred from the data.
However, data-driven learning requires data samples containing high information content to learn efficiently. For this reason, we use the \gls{sac} learning algorithm~\cite{haarnoja_soft_2019}. \gls{sac} encourages exploring new data samples where the algorithm's understanding is low, generating the necessary high quality data faster than random exploration.

In the next section, we introduce the system model of cooperative multibeam satellite communication, the applied \gls{csit} error models, typical precoding approaches, and the sum rate maximization problem. Following that, we explain the \gls{sac} method as it is used in this paper. We then apply \gls{sac} to maximize the sum rate in the presence of \gls{csit} error, and discuss \& contrast the performance against common \gls{mmse} and \gls{oma} precoding. Finally, we briefly discuss the scalability of the model case presented in this paper. For brevity, we assume prior knowledge of deep \glspl{nn}~\cite{goodfellow_deep_2020}.
 
\textit{Notations}: Lower and upper boldface letters denote vectors $\mathbf{x}$ and matrices $\mathbf{X}$ with $\mathbf{I}_N$ being an identity matrix of size $N \times N$. Transpose and Hermitian operators are indicated by $\{\cdot\}^\text{T}$ and $\{\cdot\}^\text{H}$, whereas $\circ$ is the Hadamard product. $|\cdot |$ and $\| \cdot \|$ signify the absolute value and Euclidean norm, respectively.


%
	

\section{Satellite Communication Setup \& Notations}
\label{sec:setup}

This section introduces the applied \gls{los} channel model with errors regarding the measurement of user and satellite positions. Further, we explain two common precoding techniques for satellite communications.

\subsection{System Model}
\label{sec:systemmodel}

In this paper, we consider a multi-user downlink scenario with \( \numsatellites \) \gls{leo} satellites, each equipped with a \gls{ula} of \(\numantennasper\) antennas, serving \(\numusers\) handheld users with one receive antenna each and low receive antenna gain \( \usergain \). It is assumed that all satellites are provided with the data symbols of all users and perform joint precoding, \ie all satellites perform the precoding together.
The data symbol $\datasymbol_\useridx$ of user $\useridx$ is weighted by a precoding vector $\precodingvec_\useridx \in \mathbb{C}^{\numsatellites \numantennasper \times 1}$ and transmitted over the \gls{los} channel \(\csivector_\useridx \in \mathbb{C}^{1\times \numsatellites \numantennasper}\) with \gls{awgn} $n_\useridx \sim \mathcal{CN}(0, \noisepower)$. Hence, the receive signal $y_\useridx$ follows as
\begin{align}
\label{eq:transmission}
\textstyle
    y_\useridx = \csivector_\useridx\precodingvec_\useridx \datasymbol_\useridx + \csivector_\useridx\sum^\numusers_{\otheruseridx \neq \useridx}  \precodingvec_\otheruseridx \datasymbol_l + n_\useridx .
\end{align}
The \gls{los} channel vector ~\( \csivector_{\useridx, \satidx} \in \mathbb{C}^{1\times \numantennasper} \) from satellite \(\satidx\) to user \(\useridx\) with $\csivector_\useridx = [\csivector_{\useridx,1} \dots \csivector_{\useridx, \numsatellites}]$ is given by
%
\begin{align}
\label{eq:channel_perfect}
    \csivector_{\useridx, \satidx}(\aod_{\useridx, \satidx}) = 
     \frac{\wavelength\sqrt{\usergain \satgain }}{4\pi \dist_{\useridx, \satidx}}  \text{e}^{-j \varphi_{\useridx, \satidx} }  \steeringvec_{\useridx, \satidx} ( \cos(\aod_{\useridx, \satidx})),
\end{align}
where $\satgain$ denotes the satellite antenna gain, $\wavelength$ is the wavelength and $\dist_{\useridx, \satidx}$ is the distance between the $\satidx$-th satellite and the $\useridx$-th user. The overall phase shift from satellite $\satidx$ to user $\useridx$ is given by $\varphi_{\useridx, \satidx} \in [0, 2 \pi]$. The relative phase shifts between the $\numantennasper$ antennas of a satellite $\satidx$ to user $\useridx$ are described by the steering vector $\steeringvec_{\useridx, \satidx}( \cos(\aod_{\useridx, \satidx})) \in \mathbb{C}^{1 \times N}$, where $\aod_{\useridx, \satidx}$ is the \gls{aod} from satellite $\satidx$ to user $\useridx$. The $\antidx$-th entry of the steering vector $\steeringvec_{\useridx, \satidx}(\cos(\aod_{\useridx, \satidx}))$ is given as
\begin{align}
\label{eq:steering_vec}
     \steeringentry^\antidx_{\useridx, \satidx}(\cos(\aod_{\useridx, \satidx})) = \text{e}^{-j\pi \frac{\antdist}{\wavelength} (\numantennasper +1 - 2\antidx) \cos(\aod_{\useridx, \satidx})}  ,
\end{align}
where $\antdist$ is the inter-antenna-distance between the $\numantennasper$ antennas per satellite. 
Given that the inter-antenna-distance $\antdist$ is fixed and known, the phase differences in \eqref{eq:steering_vec} between antenna $\antidx$ and $\antidx'$ are only determined by the $\useridx$\nobreakdash-th user position, reflected in the space angle $\cos(\aod_{\useridx, \satidx})$.
In practice, a precise estimation of the user's position may not be feasible due to the high velocities of \gls{leo} satellites. 
We assume that user $\useridx$'s position can be well estimated within a certain region. Therefore, we model the estimation error on the space angle $\cos(\aod_{\useridx, \satidx})$ as a uniformly distributed additive error~${\erroraod_{\useridx, \satidx} \sim \mathcal{U}(-\errorbound, + \errorbound)}$~\cite{MaikICC}.
Since the error $\erroraod_{\useridx, \satidx} $ is added on the cosine of the \glspl{aod} and given the definition of the steering vector \eqref{eq:steering_vec}, the erroneous channel vector~\( \tilde{\csivector}^{(1)}_{\useridx,\satidx}(\aod_{\useridx, \satidx}) \) can be expressed as an overall multiplicative error $\steeringvec_{\useridx, \satidx}(\erroraod_{\useridx, \satidx})\in \mathbb{C}^{1 \times \numantennasper} $ on the true channel ${\csivector}_{\useridx,\satidx}(\aod_{\useridx, \satidx})$
\begin{align}
\label{eq:error}
    \tilde{\csivector}^{(1)}_{\useridx, \satidx}( \aod_{\useridx, \satidx}, \erroraod_{\useridx,\satidx}) = \csivector_{\useridx, \satidx}( \aod_{\useridx, \satidx}) \circ \steeringvec_{\useridx, \satidx}\big(\erroraod_{\useridx, \satidx}\big) 
\end{align}
with the $\antidx$-th entry of the error vector $\steeringvec_{\useridx, \satidx}(\erroraod_{\useridx, \satidx})$
\begin{align}
\label{eq:steering_error}
     \steeringentry^\antidx_{\useridx, \satidx}(\erroraod_{\useridx, \satidx}) = \text{e}^{-j\pi \frac{\antdist}{\wavelength} (\numantennasper +1 - 2\antidx) \erroraod_{\useridx, \satidx}}  .
\end{align}

Joint precoding with multiple satellites requires tight inter-satellite synchronization. If such synchronization is imperfect, the signals from different satellites arrive with different phases. To model such synchronization mismatch, we assume an error $\errorphase_{\useridx, \satidx} \sim \mathcal{N}(0,\errorphasevariance)$ on the overall phase shift $\overallphase_{\useridx, \satidx}$. The resulting erroneous channel estimation with imperfect synchronization and position knowledge is given as
%
\begin{align}
\label{eq:error2}
    \tilde{\csivector}^{(2)}_{\useridx, \satidx}( \aod_{\useridx, \satidx}, \erroraod_{\useridx,\satidx}, \errorphase_{\useridx,\satidx} )= \text{e}^{-j \errorphase_{\useridx, \satidx}}
    \cdot
    \tilde{\csivector}^{(1)}_{\useridx, \satidx}( \aod_{\useridx, \satidx}, \erroraod_{\useridx,\satidx}).
\end{align}
In this paper, we analyze the influence of both error models, \eqref{eq:error} and \eqref{eq:error2}, on the precoding performance.
We evaluate the performance of our different precoding techniques by comparing their corresponding sum rates
\begin{align}
\label{eq:sumRate}
    \sumrate = \sum_{\useridx = 1}^{\numusers}  \log \Bigg(1+\frac{\left|\csivector_\useridx\precodingvec_{\useridx}\right|^2}{\noisepower+ \sum_{\otheruseridx \neq \useridx}^\numusers|\csivector_\useridx \precodingvec_{\otheruseridx}|^2}\Bigg) 
\end{align}
 Our goal is to maximize the sum rate~\eqref{eq:sumRate} for various satellite-to-user constellations while enhancing robustness against imperfect user position knowledge at the satellites~\eqref{eq:error} as well as imperfect satellite position knowledge~\eqref{eq:error2}. We propose to learn a robust precoding algorithm using the \gls{sac} method, to be introduced in \refsec{sec:sac}. Further, we analyze two common approaches explained in the subsequent section.

\subsection{Baseline Precoding Techniques}
\label{sec:SETUPTOPIC2}

This subsection presents two common precoding approaches for satellite downlink communications: the conventional \gls{mmse} precoder for \gls{sdma}, and an \gls{oma} approach assuming orthogonal time or frequency resources. 

\subsubsection{\gls{mmse}}
\label{sec:MMSE}
For optimal \gls{sdma} precoding, the precoding vector $\precodingvec_\useridx$ steers a beam with maximal power into the direction of the corresponding user $\useridx$ while minimizing \gls{iui}. The \gls{mmse} precoder has been proven to be a reliable precoder in this manner and is widely used \cite{windpassinger2004detection, MMSEspace}. For \gls{mmse}, the precoding matrix $\precodingmatrix^\text{MMSE} = \big[\precodingvec_1^\text{MMSE} \dots \precodingvec_\numusers^\text{MMSE}\big]$ for a given channel estimate $\tilde{\csimatrix} = [\tilde{\csivector}_1 \dots \tilde{\csivector}_\numusers]^\text{T}$ is calculated as follows
\begin{align}
	\label{eq:MMSE}
	\begin{split}
		&\precodingmatrix^\text{MMSE}= \sqrt{\frac{\transmitpower}{\text{tr} \{ {{\precodingmatrix^{\prime}}^{\text{H}}} \precodingmatrix^{\prime} \} }} \cdot \precodingmatrix^{\prime}  
\\[0.8ex]
	&\precodingmatrix^{\prime} = \Big[ \mathbf{\tilde{\csimatrix}}^\text{H} \mathbf{\tilde{\csimatrix}} + \noisepower \cdot \frac{\numusers}{\transmitpower} \cdot \mathbf{I}_{\numsatellites \numantennasper} \Big]^{-1} \mathbf{\tilde{\csimatrix}}^\text{H} 
	\end{split} ,
\end{align}
where $\transmitpower$ denotes the total amount of transmit power. In this paper, we further constrain the maximum transmit power per satellite $\transmitpower_\satidx$ to be equally distributed among all $\numsatellites$ satellites, such that $\transmitpower_\satidx \leq \transmitpower/\numsatellites$ applies. Note that the \gls{mmse} precoding approach does not necessarily maximize the sum rate $\sumrate$ \eqref{eq:sumRate}.

\subsubsection{\gls{oma}}

In contrast to \gls{sdma}, \gls{oma} uses orthogonal time or frequency resources. In this case there is no \gls{iui} between the user channels. Therefore, the optimal precoder is a \gls{mrt} precoder that solely maximizes the transmit power steered into the direction of a given user $k$
\begin{align}
    \precodingvec_\useridx^\mathrm{MRT} = \sqrt{\transmitpower} \cdot \frac{\tilde{\csivector}_\useridx^\mathrm{H}}{\|\tilde{\csivector}_\useridx\|} .
\end{align}
Here the total transmit power $\transmitpower$ is used for user $k$. Due to the absence of \gls{iui}, the sum rate for \gls{oma} calculates as
\begin{align}
\label{eq:oma}
    \sumrate^\mathrm{OMA} = \frac{1}{\numusers} {\sum_{\useridx = 1}^{\numusers}}  \log \Bigg(1+\frac{\left|\csivector_\useridx \precodingvec_{\useridx}\right|^2}{\noisepower}\Bigg) .
\end{align}
To guarantee a fair comparison between \gls{oma} and the different \gls{sdma} approaches, the overall rate is divided by the number of users $\numusers$, taking into account that in \gls{sdma} time and frequency resources are shared between all users.
The \gls{mmse} and the \gls{oma} approach will serve as baselines for our proposed learning algorithm. 

In the next section, we will discuss how to iteratively learn a robust precoding algorithm that approximately optimizes the sum rate \eqref{eq:sumRate} based on the current channel state estimate.


%
	

\section{Learning to Precode using Soft Actor-Critic}
\label{sec:sac}

In deep \gls{rl}, an agent learns to approximately maximize a performance metric by interacting with a system and adjusting its behavior according to the system feedback.
Here, the metric is the sum rate \( \reward \)~\eqref{eq:sumRate} and the behavior corresponds to the precoding algorithm.
In this section, we introduce the \acrfull{sac}~\cite{haarnoja_soft_2019} \gls{rl} method,
which is characterized by rewarding exploration where uncertainty is high, thereby generating high quality data samples.

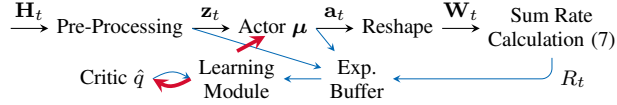
\begin{figure}[!t]
    \centering
    \begin{tikzpicture}
\tikzstyle{A0} = [-, >={stealth}, rounded corners]
\tikzstyle{A1} = [->, >={stealth}, rounded corners]
\tikzstyle{A1learn} = [->, >={stealth}, rounded corners, color=uniblue2, line width=0.1mm]
\tikzstyle{A1paramupdate} = [->, >={stealth}, rounded corners, color=unired2, line width = 0.5mm]
\tikzstyle{A2} = [<->, >={stealth}, rounded corners]
\tikzstyle{selfloop} = [looseness=4]

\tikzstyle{every node}=[font=\footnotesize]

\newcommand{\xshift}{0.5}
\newcommand{\yshift}{0.1}

\node (origin)
    at (0.0, 0.0)
    []
    {};

\node (preprocessing)
    [right= \xshift of origin]
    {Pre-Processing};

\node (actor)
    [right= \xshift of preprocessing]
    {Actor \( \actornetworkvec \)};

\node (reshapeprecoder)
    [right= \xshift of actor]
    {Reshape};

\node (commsystem)
    [right = \xshift of reshapeprecoder, align=center]
    {Sum Rate\\Calculation \refeq{eq:sumRate}};

\node (buffer)
    [below right = \yshift of actor, align=center]
    {Exp.\\Buffer};

\node (learningmodule)
    [left = \xshift of buffer, align=center]
    {Learning\\Module};

\node (critic)
    [left = \xshift of learningmodule, align=center]
    {Critic \( \criticnetworksca \)};

\draw [A1]
    (origin)
        to node [above, align=center] {\( \tilde{\csimatrix}_{\timeindex} \)}
    (preprocessing);

\draw [A1]
    (preprocessing)
        to node [above, align=center] {\( \statevec_{\timeindex} \)}
    (actor);

\draw [A1]
    (actor)
        to node [above, align=center] {\( \actionvec_{\timeindex} \)}
    (reshapeprecoder);

\draw [A1]
    (reshapeprecoder)
        to node [above, align=center] {\( \precodingmatrix_{\timeindex} \)}
    (commsystem);

\draw [A1learn]
    (commsystem.south)
        |- node [right, align=center, color=black] {\( \sumrate_{\timeindex} \)}
    (buffer);

\draw [A1learn]
    (buffer)
        to
    (learningmodule);

\draw [A1learn]
    (preprocessing.east)
        to
    (buffer);

\draw [A1learn]
    (actor.east)
        to
    (buffer);

\draw [A1learn]
    (critic.east)
        to [bend left, looseness=1.2]
    (learningmodule.west);

\draw [A1paramupdate]
    (learningmodule.north)
        to 
    ([xshift=-0.1cm, yshift=-0.1cm]actor);

\draw [A1paramupdate]
    (learningmodule.west)
        to [bend left, looseness=1.2]
    (critic.east);

\end{tikzpicture}





    \caption{\gls{sac} Precoder process flow. Top row describes inference, bottom row describes learning. Blue arrows relate to learning, red arrows show \gls{nn} parameter updates.}
    \label{fig:sacprocess}
\end{figure}

Our \gls{sac} implementation uses five components, shown in \reffig{fig:sacprocess}: 1)~a Pre-Processing step, 2)~two \glspl{cnn} \( \criticnetworksca_{1} \) and \( \criticnetworksca_{2} \), 3)~an \gls{ann} \( \actornetworkvec \), 4)~an Experience Buffer, and 5)~a Learning Module.
We denote the \gls{nn} weights as \( {\paramscritic{1}, \paramscritic{2}, \paramsactor} \) for \( \criticnetworksca_{1}, \criticnetworksca_{2}, \actornetworkvec \), respectively.
The \gls{ann} will form the precoding matrix \( \precodingmatrix \). In \gls{sac}, it produces stochastic output using the reparametrization-trick:
For the \(\actionindex\)\nobreakdash-th entry~\( \actionsca_{\actionindex} \) of an output vector \( \actionvec \), the \gls{ann} produces two outputs, one of which is considered a mean \( \actornetworksca_{\actionindex, \actornetworkmean} \) and the other a scale \( \actornetworksca_{\actionindex, \actornetworkscale} \). The outputs \( {\actionsca_{\actionindex} \sim \mathcal{N}(\actornetworksca_{\actionindex, \actornetworkmean}, \actornetworksca_{\actionindex, \actornetworkscale})}\) are then sampled from a Normal distribution.

A discrete inference and learning step \( \timeindex \) looks as follows: First, the Pre-Processing step prepares the amplitude and phase components of the complex valued erroneous \gls{csit}~\( {\tilde{\csimatrix}_{\timeindex} \in \numberscomplex^{\numusers \times  \numsatellites\numantennasper}} \) in a flat state vector~\( {\statevec_{\timeindex} \in \numbersreal^{\num{1}\times \num{2}\numsatellites\numantennasper\numusers}} \) as input for the \gls{nn}.
Based on this input vector~\( \statevec_{\timeindex} \) and its current parameters~\( \paramsactortime \), the \gls{ann}~\( {\actornetworkvec} \) will output an action vector \( {\actornetworkvec_{\paramsactortime}(\statevec_{\timeindex}) = \actionvec_{\timeindex} \in \numbersreal^{\num{1} \times \num{2}\numsatellites\numantennasper\numusers}} \). 
This real-valued vector \( \actionvec \) is then reshaped into a complex-valued precoding matrix \( {\precodingmatrix_{\timeindex} \in \numberscomplex^{\numsatellites\numantennasper \times \numusers}} \) with entries
\begin{align*}
    \precodingentry^\antidx_{\timeindex, \useridx, \satidx} = \actionvec_{t, \useridx+ \satidx+ \antidx} + j \actionvec_{\timeindex, \numantennasper+\useridx+ \satidx+ \antidx },
\end{align*}
and is normalized to the available transmit power per satellite~$\transmitpower/\numsatellites$, same as the \gls{mmse} precoder.
Subsequently, precoding is performed, the resulting sum rate~\( \reward_{\timeindex} \) \refeq{eq:sumRate} is calculated. The data sample of state~\( \statevec_{\timeindex} \), action~\( \actionvec_{\timeindex} \) and result~\( \reward_{\timeindex} \) is saved in the Experience Buffer. This concludes the inference part.

Next, to improve the quality of the \gls{ann}'s outputs, its weights~\( \paramsactortime \) must be tuned. The Actor-Critic method operates as follows: We wish to update the weights to maximize the mapping of \( (\statevec, \actionvec) \rightarrow \reward \), outputting precoding that maximizes the sum rate. This mapping is, however, unknown to the learning method.
Given the collected data samples from the Experience Buffer, the \glspl{cnn} can approximate a mapping~\( {\criticnetworksca(\statevec, \actionvec) = \rewardapprox } \). If the true mapping is approximated sufficiently well, the \gls{ann}'s weights~\( \paramsactor \) can be updated to maximize the known mappings~\( \criticnetworksca \).
Weight updates are performed once per training iteration~\( \timeindex \) on both \glspl{cnn}~\( {\criticnetworksca_{1}, \criticnetworksca_{2}} \) and \gls{ann}~\( \actornetworkvec \).
The \gls{cnn}~\( {\criticnetworksca_{1}, \criticnetworksca_{2}} \) are updated in a supervised learning manner, minimizing the squared distance
\begin{align*}
    \losscritic =
        (\criticnetworksca_{\paramscritictimegeneric}(\statevec_{\timeindex}, \actionvec_{\timeindex}) - \reward_{\timeindex})^2
    = (\rewardapprox_{\timeindex} - \reward_{\timeindex})^2
\end{align*}
between the approximation \( \hat{\reward}_{\timeindex} \) given current weights~\( \paramscritictimegeneric \) and the data sample target~\( \reward_{\timeindex} \).
We then construct an \gls{ann} loss
\begin{align*}
    \lossactorvalue = 
        -\min(
            \criticnetworksca_{1, \paramscritictime{1}}(\statevec_{\timeindex},
            \actornetworkvec_{\paramsactortime}(\statevec_{\timeindex})),
            \criticnetworksca_{2, \paramscritictime{2}}(\statevec_{\timeindex}, \actornetworkvec_{\paramsactortime}(\statevec_{\timeindex}))
        )
\end{align*}
that minimizes the negative approximate expected results~\( \hat{\reward}_{\timeindex} \), thereby maximizing the approximate expected results. Selecting conservatively from multiple, independently initialized mappings \( \criticnetworksca \) has been shown to stabilize the learning process.

This loss, however, gives no incentive for the \gls{ann} to maintain its stochastic output's scales~\( \actornetworksca_{\actionindex, \actornetworkscale} \), \ie incentive to explore new data samples where uncertain. During the learning process, insufficient exploration might lead to an incomplete data set that lacks information required to discover solutions near the global optimum.
To prevent the \gls{ann} from contracting its scales~\( \actornetworksca_{\actionindex, \actornetworkscale} \) too rapidly, the \gls{sac} method additionally adds an entropy loss
\begin{align*}
    \lossactorentropy = 
        \frac{1}{\num{2}\numsatellites\numantennasper\numusers}
        \sum_{\actionindex=1}^{\num{2}\numsatellites\numantennasper\numusers}
            \exp(\logentropyscale) \log(\pi(\actornetworksca_{\actionindex, \paramsactortime}(\statevec_{\timeindex}), \paramsactortime)),
\end{align*}
where \( \pi(\actornetworksca_{\actionindex,\paramsactortime}(\statevec_{\timeindex}), \paramsactortime) \) is the probability of \( \actornetworksca_{\actionindex, \paramsactortime}(\statevec_{\timeindex}) \) given the current weights~\( \paramsactortime \) of the \gls{ann}~\( \actornetworkvec \) and \( \logentropyscale \) is a weighting factor.
In the mean sense, this loss is directly minimized by increasing the \gls{ann} output variance and therefore represents the necessary counterweight to the first loss~\( \lossactorvalue \).
The scaling factor \( \logentropyscale \) is updated iteratively to keep the entropy roughly constant over the duration of training, \ie it is adjusted whenever the entropy falls above or below a heuristic target value. Summarily, the \gls{ann}~\( \actornetworkvec \) updates its weights~\( \paramsactor \) to minimize a composite loss \( {\lossactor = \lossactorvalue + \lossactorentropy} \).
Both \gls{ann} and \gls{cnn} parameter updates are performed using \gls{sgd}-like optimization on batches of \( \learningbatchsize \)~data samples drawn from the Experience Buffer with uniform probability.

In the following section, we evaluate specific implementation details and the performance of the learned precoder for given scenarios.


%
	

\section{Evaluation}
\label{sec:experiments}
In this section, we cover specific implementation details and compare the performance of \gls{sac}-learned precoders to the \gls{mmse} and \gls{oma} baselines.

\subsection{Implementation Details}
\label{sec:implementationdetails}
The code is implemented in Python using the TensorFlow library and is available online at \cite{gracla_code_2023}. \reftab{tab:parameters} lists the key communication and learning parameters.
For each iteration \( \timeindex \), user positions are updated around a mean inter-user-distance~${\meanuserdist = \SI{1}{\km}}$ following a uniform distribution of \( \pm \)\,\SI{30}{\meter}.
This is done to prevent the learning process from honing in on a fixed user position regardless of the information it is fed, but it also highlights the method's ability to learn a single precoding algorithm that serves different satellite and user constellations.

\begin{table}[!t]
	\renewcommand{\arraystretch}{1.3}
	\caption{Selected Parameters}
	\label{tab:parameters}
    \medskip
	\centering
    \addtolength{\tabcolsep}{-.06cm}
	\rowcolors{2}{white}{uniblue1!5} 
	\begin{tabular}{llll}
		\hline
        Noise Power $\noisepower$ & $6\text{e-}13\,\text{W}$
        &
        Transmit Power $\transmitpower$ & \SI{100}{W}
        \\
        Sat. Altitude & \SI{600}{\km} 
            &
        Ant. per Sat. $\numantennasper$ & \num{2}
        \\
        Sat. Nr. $\numsatellites$ & \num{2}
			&
		Gain per Sat. Ant $\satgain$ & \SI{14}{dBi}
        \\
        Sat. Distance & \SI{10}{km}
			&
		Gain per User $\usergain$ & \SI{0}{dBi}
		\\
        User Nr. $\numusers$ & \num{3}
                &
        Inter-Ant.-Distance $\antdist $ & $ 3 \wavelength/2$
        \\
        Wavelength $\wavelength$ & \SI{15}{cm}
            &
            Hidden Layers \( \times \) Nodes & \( 4 \times 512 \) \\
        Learning Steps \( \timeindex \) & \( 3\text{e}4 \)
            &
            Critic Learning Rate & \( 1\text{e-}5 \)
        \\
        Exp Buffer Size & \( 1\text{e}4 \)
            & Actor Learning Rate & \( 1\text{e-}6 \)
        \\
        Entropy Target~\( \logentropyscale \) & \num{1.0}
            & Data Batch Size~\( \learningbatchsize \) & \( \num{512} \)
        \\
		\hline
	\end{tabular}
\end{table}

We train three \gls{sac} precoders named \textrm{SAC1}, \textrm{SAC2}, \textrm{SAC3}. \textrm{SAC1} is trained with perfect \gls{csit}, \textrm{SAC2} is trained on error model 1 \refeq{eq:error} with a bound of \( {\errorbound = \num{0.1}} \), and \textrm{SAC3} is trained on error model 2 \refeq{eq:error2} with a bound of \( {\errorbound = \num{0.1}} \) and a scale of \( {\errorphasescale = \num{0.01}} \). 
All precoders are evaluated in terms of the achieved sum rate $\sumrate$. In cases with imperfect \gls{csit}, the sum rate $\bar{\sumrate}$ is averaged over \( \num{10000} \) Monte Carlo iterations.

\subsection{Results}
\label{sec:results}

\begin{figure}[!t]
	\centering
    \input{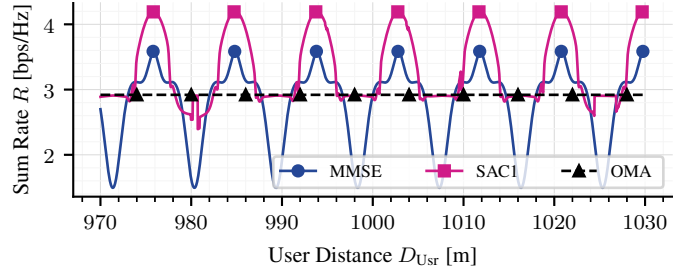}
	\caption{%
        The sum rate~\eqref{eq:sumRate} for different user distances $\userdist$ with perfect \gls{csit} clearly highlights the performance gains of the \gls{sac} precoder over \gls{mmse}, particularly when channels are highly correlated (valleys) or nearly orthogonal (peaks). The \gls{sac} precoder shows glitches, \eg around \SI{980}{\meter}; it achieves its performance despite not having reached full convergence during training.
    }
	\label{fig:distance_sweep}
\end{figure}

We first investigate the details of the precoding algorithm SAC1 for perfect \gls{csit} in \reffig{fig:distance_sweep}.
As the user distances $\userdist$ change, the user correlation in user channels alternates periodically, significantly impacting the achievable performance. The learned precoder shows gains over the \gls{mmse} precoder especially in the extreme areas, with either highly correlated or nearly orthogonal channels. When the channels are highly correlated, the learned precoder achieves its gain by focusing as much power as possible on a single user, thereby decreasing interference, but sacrificing fairness.
Crucially, the learned precoder outperforms \gls{oma} almost everywhere, demonstrating the increased spectral efficiency by frequency reuse.
In further investigations, the learned algorithms generalize well to user distances up to at least \( \pm\,\SI{300}{\meter} \) for the given scenario.
We attribute this to the learning algorithm capturing the locally periodic nature of the optimization objective during training.

\begin{figure}[!t]
	\centering
    \input{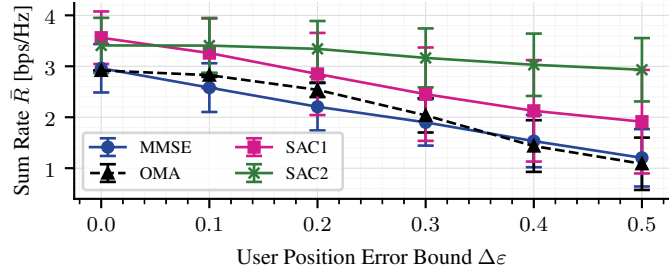}
	\caption{%
        Mean sum rate performance of different precoding approaches in the presence of an error on \gls{csit} according to the first error model \refeq{eq:error}. Error bars represent the standard deviation. Note that user distances are also varied within each evaluation, hence the variance even at zero error.
    }
	\label{fig:error_sweep_1}
\end{figure}

\reffig{fig:error_sweep_1} compares SAC1 and SAC2 to \gls{mmse} and \gls{oma} precoding with unreliable user position measurements \eqref{eq:error} for increasing error bound $\errorbound$.
As expected, since \gls{mmse} is not sum rate optimal, SAC1 achieves the best sum rate performance with perfect \gls{csit}, with SAC2 slightly behind and both outperforming the baselines \gls{mmse} and \gls{oma}. With increasingly unreliable user position estimates, the performance of all four precoders degrades. However, SAC2, which already encountered unreliable information during training, shows resilience even at very high values of \( \errorbound \).

\begin{figure}[!t]
	\centering
    \input{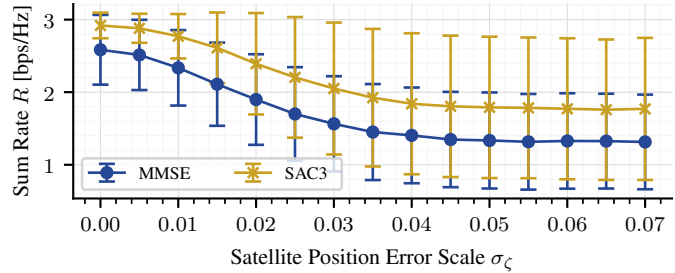}
	\caption{%
        Mean sum rate performance on the second error model, \refeq{eq:error2}. We fixate the user position error bound to \( \errorbound = 0.1 \) and sweep over the satellite position error scale~\( \errorphasescale \). Again, the learned algorithm achieves higher sum rates than the \gls{mmse} precoder, as well as slightly increased robustness at higher errors.
    }
	\label{fig:error_sweep_2}
\end{figure}

We then confirm in \reffig{fig:error_sweep_2} that the same learning approach also works for imperfections of satellite positions~\refeq{eq:error2}, demonstrating the adaptability of model-free data-driven optimization. 

%
	

\section{Conclusions}
\label{sec:conclusions}

We have applied the \gls{sac} \gls{rl} method to the problem of sum rate optimal precoding for cooperative multibeam satellite communications in the presence of erroneous channel knowledge. Our results show that we are able to learn both effective and robust precoding algorithms with no assumptions on the underlying error model.
We therefore expect this approach to scale well on real life satellites, particularly with sights on dedicated \gls{ml} hardware being deployed on future satellites~\cite{giuffrida_cloudscout_2020}.


%
	%
	%
	%
	\bibliographystyle{ref/IEEEbib}%
    \bibliography{%
        ref/references
    }%

\begin{thebibliography}{10}

\bibitem{3GPP.TR.38.863}
{3GPP TR 38.863},
\newblock ``Technical specification group radio access network; solutions for
  nr to support non-terrestrial networks ({NTN}): Non-terrestrial networks
  ({NTN}) related {RF} and co-existence aspects (release 17),'' Sep. 2022.

\bibitem{Leyva-Mayorga2020}
Israel Leyva-Mayorga, Beatriz Soret, Maik Röper, Dirk Wübben, Bho Matthiesen,
  Armin Dekorsy, and Petar Popovski,
\newblock ``{LEO} small-satellite constellations for {5G} and beyond-{5G}
  communications,''
\newblock {\em IEEE Access}, vol. 8, pp. 184955--184964, 2020.

\bibitem{Qu}
Zhicheng Qu, Gengxin Zhang, Haotong Cao, and Jidong Xie,
\newblock ``{LEO Satellite Constellation for Internet of Things},''
\newblock {\em IEEE Access}, vol. 5, pp. 18391--18401, 2017.

\bibitem{vazquez2018precoding}
Miguel~{\'A}ngel V{\'a}zquez, MR~Bhavani Shankar, Charilaos~I Kourogiorgas,
  Pantelis-Daniel Arapoglou, Vincenzo Icolari, Symeon Chatzinotas, Athanasios~D
  Panagopoulos, and Ana~I P{\'e}rez-Neira,
\newblock ``Precoding, scheduling, and link adaptation in mobile interactive
  multibeam satellite systems,''
\newblock {\em IEEE Journal on Selected Areas in Communications}, vol. 36, no.
  5, pp. 971--980, 2018.

\bibitem{MaikBeamspace}
Maik Röper, Bho Matthiesen, Dirk Wübben, Petar Popovski, and Armin Dekorsy,
\newblock ``{Beamspace MIMO for Satellite Swarms},''
\newblock in {\em Proc. IEEE Wireless Commun. Netw. Conf. (WCNC)}, 2022, pp.
  1307--1312.

\bibitem{liu2022robust}
Yanhao Liu, Yibiao Wang, Jue Wang, Li~You, Wenjin Wang, and Xiqi Gao,
\newblock ``Robust downlink precoding for leo satellite systems with
  per-antenna power constraints,''
\newblock {\em IEEE Transactions on Vehicular Technology}, vol. 71, no. 10, pp.
  10694--10711, 2022.

\bibitem{mnih2013playing}
Volodymyr Mnih, Koray Kavukcuoglu, David Silver, Alex Graves, Ioannis
  Antonoglou, Daan Wierstra, and Martin Riedmiller,
\newblock ``{Playing Atari with Deep Reinforcement Learning},''
\newblock {\em arXiv:1312.5602}, 2013.

\bibitem{dahrouj_overview_2021}
Hayssam Dahrouj, Rawan Alghamdi, Hibatallah Alwazani, Sarah Bahanshal,
  Alaa~Alameer Ahmad, Alice Faisal, Rahaf Shalabi, Reem Alhadrami, Abdulhamit
  Subasi, Malak~T. Al-Nory, Omar Kittaneh, and Jeff~S. Shamma,
\newblock ``An {Overview} of {Machine} {Learning}-{Based} {Techniques} for
  {Solving} {Optimization} {Problems} in {Communications} and {Signal}
  {Processing},''
\newblock {\em IEEE Access}, vol. 9, pp. 74908--74938, 2021.

\bibitem{zhang_data_2022}
Shaoqing Zhang, Jindan Xu, Wei Xu, Ning Wang, Derrick Wing~Kwan Ng, and Xiaohu
  You,
\newblock ``Data augmentation empowered neural precoding for multiuser mimo
  with mmse model,''
\newblock {\em IEEE Communications Letters}, vol. 26, no. 5, pp. 1037--1041,
  2022.

\bibitem{lee_deep_2020}
Heunchul Lee, Maksym Girnyk, and Jaeseong Jeong,
\newblock ``Deep reinforcement learning approach to {MIMO} precoding problem:
  {Optimality} and {Robustness},'' June 2020,
\newblock arXiv:2006.16646.

\bibitem{sohrabi_robust_2020}
Foad Sohrabi, Hei~Victor Cheng, and Wei Yu,
\newblock ``Robust {Symbol}-{Level} {Precoding} {Via} {Autoencoder}-{Based}
  {Deep} {Learning},''
\newblock in {\em {ICASSP} 2020 - 2020 {IEEE} {International} {Conference} on
  {Acoustics}, {Speech} and {Signal} {Processing} ({ICASSP})}, Barcelona,
  Spain, May 2020, pp. 8951--8955, IEEE.

\bibitem{haarnoja_soft_2019}
Tuomas Haarnoja, Aurick Zhou, Kristian Hartikainen, George Tucker, Sehoon Ha,
  Jie Tan, Vikash Kumar, Henry Zhu, Abhishek Gupta, Pieter Abbeel, and Sergey
  Levine,
\newblock ``Soft {Actor}-{Critic} {Algorithms} and {Applications},''
\newblock {\em arXiv:1812.05905}, Jan. 2019.

\bibitem{goodfellow_deep_2020}
Ian Goodfellow, Yoshua Bengio, and Aaron Courville,
\newblock {\em Deep {Learning}},
\newblock MIT Press, 2020.

\bibitem{MaikICC}
Maik Röper, Bho Matthiesen, Dirk Wübben, Petar Popovski, and Armin Dekorsy,
\newblock ``Robust precoding via characteristic functions for vsat to
  multi-satellite uplink transmission,'' 2023.

\bibitem{windpassinger2004detection}
Christoph Windpassinger,
\newblock {\em {Detection and precoding for multiple input multiple output
  channels}},
\newblock Ph.D. thesis, Friedrich-Alexander-Universit{\"a}t
  Erlangen-N{\"u}rnberg (FAU), 2004.

\bibitem{MMSEspace}
Symeon Chatzinotas, Gan Zheng, and Björn Ottersten,
\newblock ``{Energy-efficient MMSE beamforming and power allocation in
  multibeam satellite systems},''
\newblock in {\em 2011 Conference Record of the Forty Fifth Asilomar Conference
  on Signals, Systems and Computers (ASILOMAR)}, 2011, pp. 1081--1085.

\bibitem{gracla_code_2023}
Steffen Gracla and Alea Schröder,
\newblock ``{Learning Beamforming},''
  \url{https://github.com/Steffengra/2302_learning_beamforming_code}, 2023.

\bibitem{giuffrida_cloudscout_2020}
Gianluca Giuffrida, Lorenzo Diana, Francesco de~Gioia, Gionata Benelli,
  Gabriele Meoni, Massimiliano Donati, and Luca Fanucci,
\newblock ``{CloudScout}: {A} {Deep} {Neural} {Network} for {On}-{Board}
  {Cloud} {Detection} on {Hyperspectral} {Images},''
\newblock {\em Remote Sensing}, vol. 12, no. 14, pp. 2205, July 2020.

\end{thebibliography}
	%
	%
	%
\end{document}